\documentclass[twocolumn,showpacs,prd,aps,tightenlines,floatfix,final]{revtex4}

\usepackage{graphics,graphicx}
  \DeclareGraphicsExtensions{.eps}   
\usepackage{color}
\usepackage{amsmath,amssymb}

\usepackage{array}
\newcolumntype{x}[1]{%
>{\centering\hspace{0pt}}p{#1}}%

\usepackage{warpcol}
\newcolumntype{L}[1]{>{\pcolbegin{r}{#1}}l<{\pcolend}}
\newcolumntype{R}[1]{>{\pcolbegin{r}{#1}}r<{\pcolend}}

\newcommand{\TG}{\tilde{\Gamma}}
\newcommand{\psimm}{\psi^{-2}}
\newcommand{\er}{\eta(\vec{r})}

\newcommand{\order}[1]{$#1^{{\mbox{\small th}}}$}
\newcommand{\gt}{\tilde{\gamma}}

\begin{document}

\title{
Dynamical shift condition for unequal mass black hole binaries
}

\author{Doreen~M\"{u}ller, Jason Grigsby, Bernd~Br\"{u}gmann}

\affiliation{Theoretical Physics Institute, 
University of Jena, 07743 Jena, Germany}

\date{\today}

\begin{abstract}
Certain numerical frameworks used for the evolution of binary black
holes make use of a gamma driver, which includes a damping factor. Such
simulations typically use a constant value for damping. However, it
has been found that very specific values of the damping factor are
needed for the calculation of unequal mass binaries. We examine
carefully the role this damping plays, and provide two explicit,
non-constant forms for the damping to be used with mass-ratios further
from one. Our analysis of the resultant waveforms compares well
against the constant damping case.
\end{abstract}


\maketitle

%%%%%%%%%%%%%%%%%%%%%%%%%%%%%%%%%%%%%%%%%%%%%%%%%%%%%%%%%%%%%%%%%%%%%%%%%%%%%%%%
\section{Introduction}
\label{sec:Introduction}

The ability to simulate the final inspiral, merger, and ring-down of
black hole binaries with numerical
relativity~\cite{Pre05,CamLouMar05,BakCenCho05b} plays a key role
in understanding a source of gravitational waves that may one day be
observed with gravitational wave detectors.  While initial simulations
focused on binaries of equal-mass, zero spin, and quasi-circular
inspirals, there currently is a large effort to explore the parameter
space of binaries, e.g.~\cite{Pre07,Han09,Spe09,AylBakBog09}.  A key
part of studying the parameter space is to simulate binaries with
intermediate mass-ratios.

To date, the mass ratio furthest from equal masses that has been
numerically simulated is 10:1 \cite{GonSpeBru08,LouNakZlo10}. These
simulations use the Baumgarte-Shapiro-Shibata-Nakamura (BSSN)
formulation \cite{ShiNak95,BauSha98,NakOohKoj87} with 1+log slicing,
and the $\TG$ driver condition for the shift \cite{AlcBru00,AlcBruPol01a}.
In~\cite{GonSpeBru08}, it was noted that the stability of the
simulation is sensitive to the damping factor, $\eta$, used in the $\TG$
driver condition,
\begin{equation}
  \partial_{0}^2 \beta^i = \frac{3}{4} \partial_0 \TG^i - \eta
\partial_0\beta^i.
  \label{eq:gammadriver}
\end{equation}
Here, $\beta^i$ is the shift vector describing how the coordinates move
inside the spatial slices, $\partial_0 \equiv \partial_t -
\beta^i\partial_i$, and $\TG^i$ is the contraction of the Christoffel
symbol, $\TG^i_{jk}$, with the conformal metric,
$\tilde{\gamma}^{jk}$.

The standard choice for $\eta$ is to set it to a constant value, which
works well even for the most demanding simulations as long as the
mass ratio is sufficiently close to unity. 
In binary simulations, a typical choice is a constant value of about
$2/M$, with $M$ the total mass of the system.  This choice, however,
leads to instabilities for the mass ratio 10:1 simulation
\cite{GonSpeBru08}, although stability was obtained for $\eta=1.375/M$.
The value of $\eta$ is chosen to damp an outgoing change in the shift
while still yielding stable evolutions. As we will show, if $\eta$ is
too small, there are unwanted oscillations, and values that are too
large lead to instabilities.
By itself, this observation is not new, see e.g.
\cite{BruGonHan06,Spe06,MetBakKop06,Sch10}.
The key issue for unequal masses is that, as evident from
(\ref{eq:gammadriver}), the damping factor $\eta$ has units of inverse
mass, $1/M$. Therefore, the interval of suitable values for $\eta$
depends on the mass of the black holes.
For unequal masses, a constant $\eta$ cannot equally well accommodate
both black holes. A constant damping parameter implies that the
effective damping near each black hole is asymmetric since the damping
parameter has dimensions $1/M$. For large mass ratios, this asymmetry
in the grid can be large enough to lead to a failure of the
simulations because the damping may become too large or too small for
one of the black holes.  To cure this problem, we need a
position-dependent damping parameter that adapts to the local mass. In
particular, we want it to vary such that, in the vicinity of the
\order{i} puncture with mass $M_i$, its value approaches $1/M_i$.

A position-dependent $\eta$ was already considered when the $\TG$
driver condition was introduced
\cite{AlcBruDie02,AlcBruDie04,ZloBakCam05,DieHerPol05,GonSpeBru08},
but such constructions were not pursued further because for moderate
mass ratios a constant $\eta$ works well.  Recently, we revived the
idea of a non-constant $\eta$ for moving puncture evolutions in order
to remove the limitations of a constant $\eta$ for large mass ratios.
In \cite{MueBru09}, we constructed a position-dependent $\eta$ using
the the conformal factor, $\psi$, which carries information both about
the location of the black holes, and about the local puncture
mass. The form of $\eta$ was chosen to have proper fall-off rates both
at the punctures and at large distance from the binary. In
\cite{LouNakZlo10}, this approach was used successfully for mass ratio
10:1.  (We note in passing that damping is useful in other gauges as
well, e.g.\ in~\cite{LinSzi09} the modified harmonic gauge condition
includes position-dependent damping by use of the lapse function.)

In the present work, we examine one potential short-coming of the
choice of \cite{MueBru09}, which leads us to suggest an alternative
type of position-dependent $\eta$. Using \cite{MueBru09}, we find
large fluctuations in the values of that $\eta$, and this might lead
to instabilities in the simulation of larger mass-ratio binary black
holes. To address this, we have tested two new explicit formulas for
the damping factor designed to have predictable behavior throughout
the domain of computation. We find the new formulas to produce only
small changes in the waveforms that diminish with resolution, and
there is a great deal of freedom in the implementation.  Independently
of our discussion here, in \cite{Sch10} the stability issues for large
$\eta$ are explained, and a non-constant $\eta$ is suggested (although
not yet explored in actual simulations), that, in its explicit
coordinate dependence, is similar to one of our suggestions.

The paper is organized as follows. We first describe the reasons
for the damping factor and some of the reasons for limiting its value
in Sec~\ref{sec:Motivation}. In Sec.~\ref{sec:forms_of_eta}, we
discuss some previous forms of $\eta$ that have been used.  We 
also present two new definitions and why we investigated them. In
Sec.~\ref{sec:Results}, we find that these new definitions agree well
with the use of constant $\eta$ in the extracted gravitational waves
for mass ratios up to 4:1. Finally, in Sec.~\ref{sec:Discussion}, we
discuss further implications of this work.

%%%%%%%%%%%%%%%%%%%%%%%%%%%%%%%%%%%%%%%%%%%%%%%%%%%%%%%%%%%%%%%%%%%%%%%%%%%%%%%%

\section{Motivation}\label{sec:Motivation}
In order to define a position-dependent form for $\eta$, it is
important to determine what this damping parameter accomplishes
in numerical simulations.  For this reason, we examine the effects of
running different simulations while varying $\eta$ between runs.
First we use evolutions of single non-spinning black-holes to identify
the key physical changes.  Then we examine equal-mass binaries to
determine specific values desired in $\eta$ at both large and small
radial coordinates.

\subsection{Numerics}\label{sec:numerics}
For all the work in this paper, we have used the BAM computer code
described in \cite{BruGonHan06,HusGonHan07,BruTicJan03}. It
uses the BSSN formalism with 1+log slicing and $\TG$ driver condition
in the moving puncture framework
\cite{CamLouMar05,BakCenCho05}. Puncture initial data \cite{BraBru97}
with Bowen-York extrinsic curvature \cite{BowYor80} have been used
throughout this work, solving the Hamiltonian constraint with the
spectral solver described in \cite{AnsBruTic04}.  For binaries,
parameters were chosen using \cite{WalBruMue09} to obtain
quasi-circular orbits, while the parameters for single black holes were
chosen directly.  We extract waves via the Newman-Penrose scalar
$\Psi_4$. The wave extraction procedure is described in detail in
\cite{BruGonHan06}.  We perform a mode decomposition using
spin-weighted spherical harmonics with spin weight $-2$,
$Y_{lm}^{-2}$, as basis functions and calculate the scalar product
\begin{equation}
 \label{eq:modes}
 \Psi_4^{lm} = \left( Y_{lm}^{-2}, \Psi_4  \right) = \int_0^{2\pi} \int_0^{\pi}
\sin \theta \mbox{d}\theta \mbox{d}\varphi\, \overline{Y_{lm}^{-2} } \Psi_4.
\end{equation}
We further split $\Psi_4^{lm}$ into mode amplitude $A_{lm}$ and phase
$\phi_{lm}$ in order to cleanly separate effects in these components,
$r_{\rm ex}\cdot\Psi_4^{lm} = A_{lm}\mbox{e}^{i \phi_{lm}}$. In this
paper, we focus on one of the most dominant modes, the $l=m=2$ mode,
and report results for this mode unless stated otherwise. The
extraction radius used here is $r_{\rm ex}=90\,M$.

\subsection{\label{sec:single_hole}Single, non-spinning puncture with
constant damping}

The damping factor, $\eta$, in Eq.~(\ref{eq:gammadriver}), is included
to reduce dynamics in the gauge during the evolution.  To examine the
problem brought up in the introduction, we compare results of a
single, non-spinning puncture with mass $M$.  We use a Courant factor
of $0.5$ and 9 refinement levels centered around the puncture. The
resolution on the finest grid is $0.025\,M$, and the outer boundary is
situated at $256\,M$. Varying the damping constant between $0.0/M$ and
$4.5/M$, two main observations can be made. First, as designed, a
non-zero $\eta$ attenuates emerging gauge waves efficiently. Second,
an instability develops for values of $\eta$ that are too large.

\begin{figure}[ht]
\includegraphics[width=\linewidth,clip]{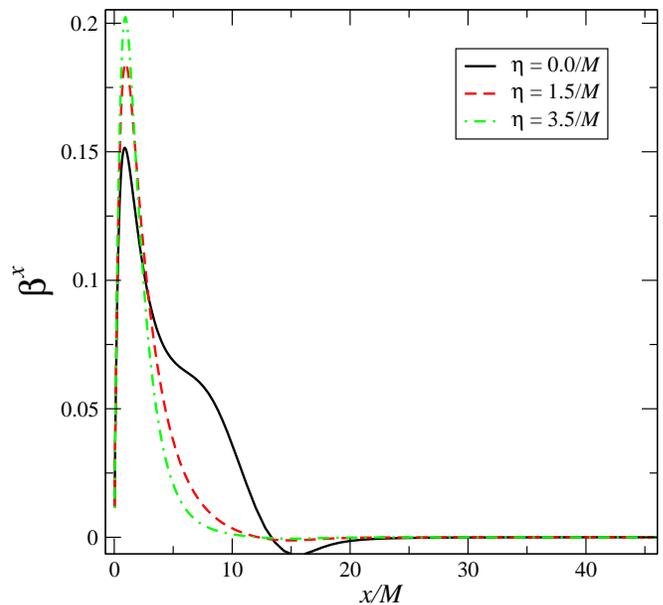}
\caption{The $x$-component of the shift, $\beta^x$, for a single
          non-spinning puncture of mass $M$ at time $t = 15.2\,M$. The three
          lines were taken for different values of the damping factor
          $\eta$.  The solid  line (black) is for $\eta = 0.0/M$.  The dashed
          line (red) is for $\eta = 1.5/M$ and the dotted-dashed line (green)
          is for $\eta = 3.5/M$. This shows the beginning of a pulse
          in $\beta^x$ for smaller values of $\eta$.
        }
        \label{fig:SPbetax1_t15.2}
\end{figure}

\begin{figure}[ht]
\includegraphics[width=\linewidth,clip]{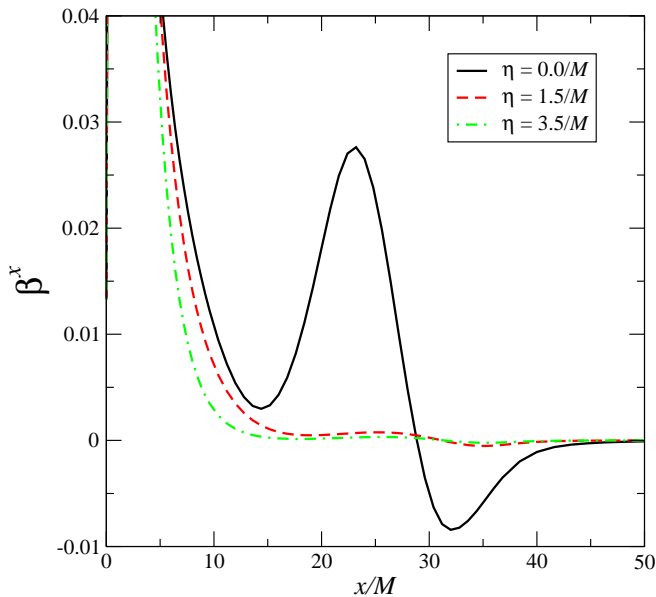}
\caption{The $x$-component of the shift, $\beta^x$, for a single
          non-spinning puncture of mass $M$ at time $t = 30.4\,M$. The three
          lines were taken for different values of the damping factor
          $\eta$ with the same line type and color scheme as in
          Fig.~\ref{fig:SPbetax1_t15.2}. Here it is clear a pulse radiates
          outward in the shift with smaller values of $\eta$. 
          }
        \label{fig:SPbetax1_t30.4}
\end{figure}

Figs.~\ref{fig:SPbetax1_t15.2}, and \ref{fig:SPbetax1_t30.4}
illustrate the first observation. Both figures show the $x$-component
of the shift along the $x$-coordinate using $\eta\in\{0.0/M, 1.5/M,
3.5/M\}$.
Apart from the usual shift profile, Fig.~\ref{fig:SPbetax1_t15.2}
shows the beginnings of a pulse in the $\eta = 0.0/M$ case (solid
line) at $x\approx 10\,M$ after $15.2\,M$ of evolution.  Examining
Fig.~\ref{fig:SPbetax1_t30.4}, where we zoom in at a later time,
$t=30.4\,M$, one can see that the pulse has started to travel further
out (solid line). Looking carefully, one can also see a much smaller
pulse in the $\eta = 1.5/M$ line (dashed).  Lastly, by examination,
one can find almost no traveling pulse in the $\eta = 3.5/M$ curve
(dotted-dashed line).  The observed pulse in the shift travels to
regions far away from the black hole and effects the gauge of distant
observers.  This might have undesirable implications for the value of
such numerical data when trying to understand astrophysical sources.
\begin{figure}[ht]
  \centering
  \includegraphics[width=\linewidth,clip]{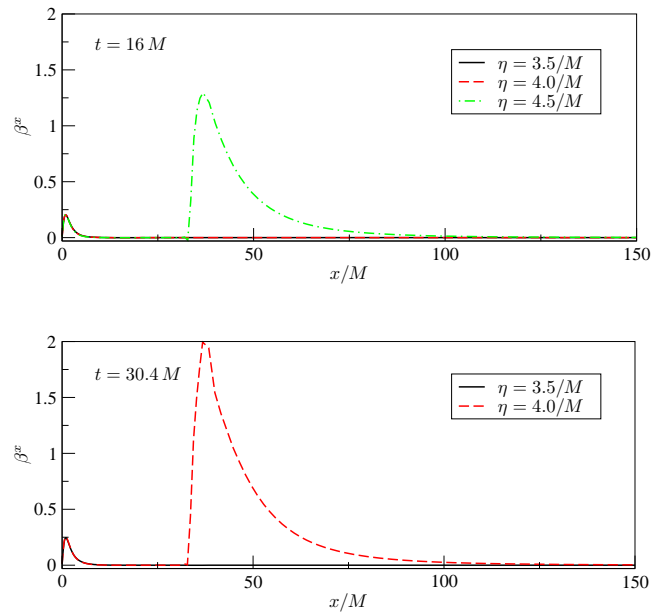}
  \caption{The $x$-component of the shift vector in the $x$-direction
    for a single non-spinning puncture of mass $M$ at times
    $t=16.0\,M$ and $t=30.4\,M$. The three different lines mark three
    values of the damping constant $\eta$. The solid line (black) is
    for $\eta=3.5/M$, the dashed line (red) for $\eta=4.0/M$ and the
    dotted-dashed line (green) for $\eta=4.5/M$. At $t=16\,M$, the
    simulation using $\eta=4.5/M$ develops an instability in the shift
    vector and fails soon afterward, the same happens for
    $\eta=4.0/M$ at $t=30.4\,M$.  In the simulation using $\eta=3.5$,
    no such instability develops (not shown).}
  \label{fig:betax_fail}
\end{figure}

For values of $\eta$ larger than $3.5/M$, an instability arises in the
shift at larger radius. Fig.~\ref{fig:betax_fail} shows the
$x$-component of the shift vector using damping constants $\eta=3.5/M$
(solid line), $\eta=4.0/M$ (dashed line) and $\eta=4.5/M$
(dotted-dashed line). The plots show an instability in simulations
with $\eta>3.5/M$ developing in $\beta^i$, which eventually leads to a
failure of the simulations. Contrary to this, the simulation using
$\eta=3.5/M$ does not show this shift related instability. In test
runs we found that by decreasing the Courant factor used, we could
increase the value of the damping factor and still get stable
evolutions. This agrees with~\cite{Sch10} where it was shown that the
gamma driver possesses the stiff property, which limits the size of
the time-step in numerical integration based on the value of the
damping.

Figures~\ref{fig:SPbetax1_t15.2},~\ref{fig:SPbetax1_t30.4},
and~\ref{fig:betax_fail} make clear how the choice of the damping
factor affects the behavior of the simulations. The value we choose
for $\eta$ should be non-zero and not larger than $3.5/M$ to allow for
effective damping and stable simulations. The exact cutoff value
between stable and unstable simulations is not relevant here since the
position dependent form we develop in Sec.~\ref{sec:forms_of_eta}
gives us the flexibility we need to obtain stable simulations.

\subsection{\label{sec:embinary}Equal mass binary with constant damping}

To examine the effect of $\eta$ on the extraction of gravitational
waves, we compare the results from simulations of an equal mass binary
with total mass $M$ in quasi-circular orbits with initial separation
$D=10\,M$, using $\eta \in \{0.0/M, 0.5/M, 2.0/M\}$.  Again, the
Courant factor is chosen to be $0.5$ and we use, in the terminology
of~\cite{BruGonHan06}, the grid configuration
$\chi[6\times56:5\times112:6]$ with a finest resolution of
$0.013\,M$. Here, the extraction radius $r_{\rm ex}$ is chosen to be
$90\,M$.

\begin{figure}
    \centering
    \includegraphics[width=\linewidth,clip]{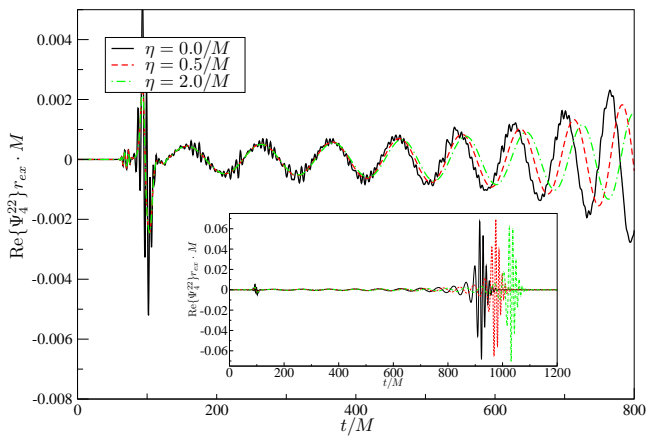}
    \caption{Real part of the 22-mode of $\Psi_4$ over time for equal mass
            simulations using different values for $\eta$. The inset shows
            the full waveform until ringdown. The solid curves (black) are
            for $\eta=0$, the dashed curve (red) mark $\eta=0.5/M$ and
            the dotted-dashed curves (green) are for $\eta=2.0/M$.
            Without damping in the shift, the extracted waves are
            noisy at times when the amplitude is still small (black, solid
            curve).
    }
    \label{fig:emexx_rpsi4}
  \end{figure}

For vanishing $\eta$, we find a lot of noise in the the real part of
the 22-mode of $r_{\rm ex}\Psi_4$, shown in the solid curve of
Fig.~\ref{fig:emexx_rpsi4}. A small, but non-vanishing $\eta$ suffices
to suppress this noise, as seen in the dashed curve of this figure.
The dotted-dashed curve in this plot is the result for using the value
$\eta=2.0/M$. We see a difference in time between peak amplitudes of
the three curves due to the change of coordinates that the alternation
of $\eta$ introduces. We did, however, find that by decreasing the
Courant factor those differences between peak amplitudes summarily
decreased.

\begin{figure}
    \centering
    \includegraphics[width=\linewidth,clip]{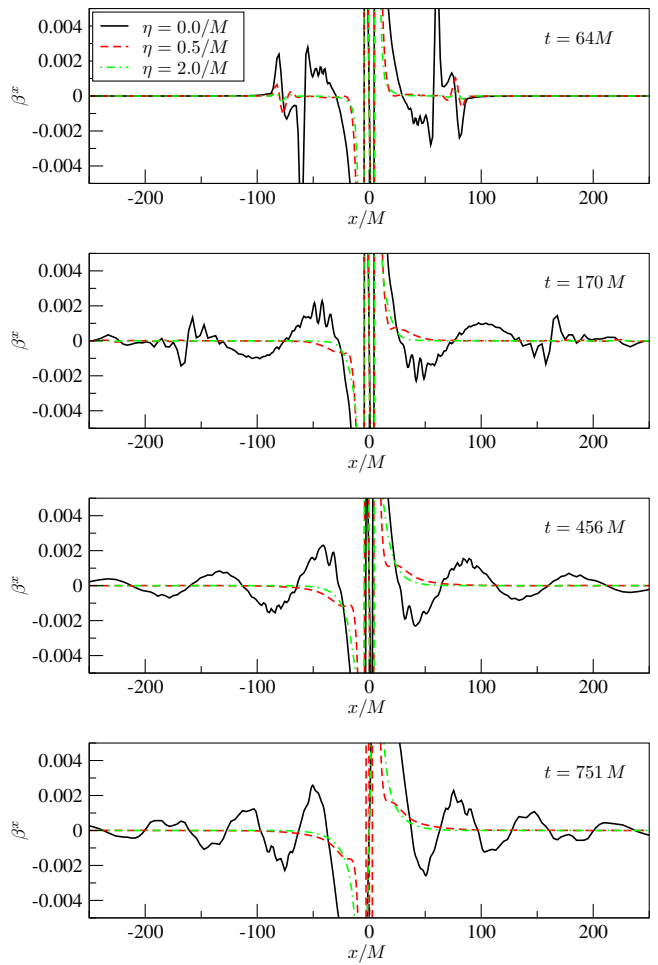}
    \caption{$x$-component of the shift vector, $\beta^x$, for three different
            choices of $\eta$ at four different times during the simulation.
            The physical system is the same as in Fig.~\ref{fig:emexx_rpsi4}.
            The merger takes place  at approximately $t=1000\,M$. In the
            $\eta=0/M$ case (black, solid curve), the shift vector is not
            damped and therefore, a pulse travels outwards and distorts
            the shift over the whole grid. The amplitude of this pulse is
            considerably damped when using a non-vanishing $\eta$ and
            therefore the distortions are reduced. For $\eta=0.5/M$ (red,
            dashed curve), there are still small bumps traveling out which are
            reduced by using $\eta=2/M$ (green, dotted-dashed curve).
            }
    \label{fig:emexx_betax}
  \end{figure}

To understand the noise in the waves for $\eta=0.0/M$, we look at the
shift vector at different times. The first panel of
Fig.~\ref{fig:emexx_betax} shows the $x$-component of the shift over
$x$, again for $\eta \in \{0.0/M, 0.5/M, 2.0/M\}$, shortly after the
beginning of the simulation. The fourth panel shows the same at a time
shortly before the merger, and the two panels in the middle represent
intermediate times. We see clear gauge pulses in the earliest time
panel for all three curves. We also observe the amplitude of this
pulse decreasing with increasing $\eta$. As time goes on, the gauge
pulse travels outwards as in the case for a single puncture in
section~\ref{sec:single_hole}.  For vanishing $\eta$ (solid line), the
shift becomes more and more distorted, and the distortions do not leave
the grid. For non-zero $\eta$, the amplitude of the gauge pulse
decreases when traveling outwards, and the shape of $\beta^x$ is not
distorted. There is, compared to $\eta=2.0/M$ (dotted-dashed line),
only a small bump left in the $\eta=0.5/M$ case (dashed line), that
changes its shape slightly during the simulation, but does not travel
to large distances from the punctures. The coordinates are disturbed
in the case where no damping is used, and thus the noise in $r_{\rm ex}
Re\{\Psi_4^{22}\}$ is not surprising.

In this series, using a Courant factor of 0.5, we only obtained stable
evolutions for $\eta<3.5/M$ which agrees with the limits found in
section~\ref{sec:single_hole}. If we chose the value of $\eta$ too large, the
same kind of instability in the shift vector we found there develops in the
equal mass case and the simulations fail. The
failure occurs relatively early, before $50\,M$ of evolution time, whereas the
stable runs lasted about $1200\,M$, including merger and ringdown (we stopped
the runs after ringdown).

%%%%%%%%%%%%%%%%%%%%%%%%%%%%%%%%%%%%%%%%%%%%%%%%%%%%%%%%%%%%%%%%%%%%%%%%%%%%%%%%
\section{Position-dependent forms of $\eta$}\label{sec:forms_of_eta}

In section~\ref{sec:single_hole}, we saw that a sufficient level of
damping is needed to limit gauge dynamics, and too much damping
can lead to numerical instabilities.  In section~\ref{sec:embinary},
we saw the positive effect that sufficient damping has on the resultant
waveform for equal mass binaries. While we still need damping in the
gamma driver in the unequal mass case, a constant value may not
fulfill the requirements of limiting gauge dynamics and permitting
stable evolutions. Rather, we need a definition for the damping that
adjusts the value to the local mass-scale.

We will examine definitions, that naturally track the position, and
mass of the individual black holes.  The choice of $\eta$ should
provide a reasonable value both near the individual black holes, and
at large distance from the binary.  We will start by examining some
previous work, that has used non-constant forms of the damping
parameter, and why it may be necessary to use other formulas.  We will
then present the two new formulas for $\eta$, which we designed for
this work.

\subsection{Previous dynamic damping parameters}\label{sec:Prev}
%%------------------------------------------------------------------------------
A position dependent damping was introduced some years ago by the
authors of \cite{AlcBruDie04}, and was later used in
\cite{ZloBakCam05}. That formula reads
\begin{equation}
  \eta = \eta_{\rm punc} - \frac{\eta_{\rm punc} -
    \eta_\infty}{1    + (\psi -1)^2}
  \label{eq:etaAlcubierre}
\end{equation}
with $\eta_{\rm punc}$, $\eta_\infty$ being constants, and assuming
$\psi = 1 + M_1/(2r_1) + M_2/(2r_2)$ ($r_i$ is the distance to the
\order{i} puncture). This formula was used to damp gauge dynamics
while using excision for equal-mass head-on collisions.  It has since
been found that using the moving puncture framework allows for
constant damping in the approximately equal mass case.  We are looking
for a formula which is suitable for the quasicircular inspiral of
intermediate mass-ratio binaries.

Previously~\cite{MueBru09}, we used the formula 
  \begin{equation}
    \er = \hat{R}_0 \frac{\sqrt{\gt^{ij}\partial_i\psimm\partial_j\psimm}} 
    {(1 - \psimm)^2},
    \label{eq:etapsimm}
  \end{equation}
\begin{figure}
 \centering
 \includegraphics[width=\linewidth,clip]{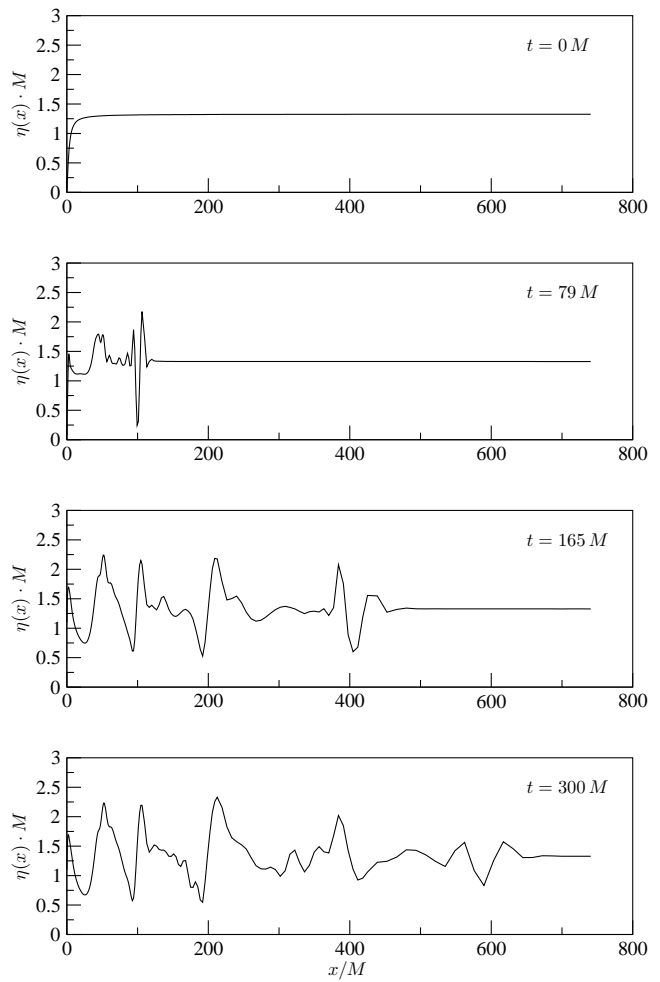}
 \caption{Damping factor, $\eta$, along the $x$-axis using
   Eq.~(\ref{eq:etapsimm}). The simulated configuration is an equal mass
   binary with initial separation $D=10\,M$ and orbits lying in the
   $(x,y)$-plane. Shown are four different times during the
   simulation.}
 \label{fig:oldeta_waves}
\end{figure}
for determining a position dependent damping coefficient instead of
using a constant $\eta$. With $\hat{R}_0$ taken to be a unitless
constant, it can be seen that Eq.~(\ref{eq:etapsimm}) has units of
inverse mass.  The dependence on the BSSN variable, $\psi$, naturally
tracks the position, and mass of the black holes. The application of
Eq.~(\ref{eq:etapsimm}) gave good values for the damping both at the
punctures, and at the outer boundary, and was even found to somewhat
decrease the grid-size of the larger black hole. The latter point
could have positive effects on how the individual black holes are
resolved on the numerical grid. It even had the additional effect of
keeping the horizon shapes roughly circular, even close to merger -
something that doesn't hold in the constant $\eta$ case. Most
importantly, the simulations remained stable, without significantly
changing the gravitational waves.  The formula was later used
successfully for the 10:1 mass-ratio in \cite{LouNakZlo10}.

Despite all this, Eq.~(\ref{eq:etapsimm}) provides reason for
concern. Fig.~\ref{fig:oldeta_waves} shows the form of $\eta$ using
Eq.~(\ref{eq:etapsimm}) for a non-spinning binary of equal mass in
quasicircular orbits starting at a separation of $D=10\,M$ at four
different times in the simulation. As can be seen, noise travels out
from the origin as time progresses. This leaves steady features on the
form of $\eta$ which could spike to higher and lower values than the range
determined in Sec.~\ref{sec:single_hole}. Additionally, these
sharp features may lead to unpredictable coordinate drifts, and could,
in some cases, affect the long-term stability of the simulation.

\begin{figure}[ht]
  \centering
  \includegraphics[width=\linewidth,clip]{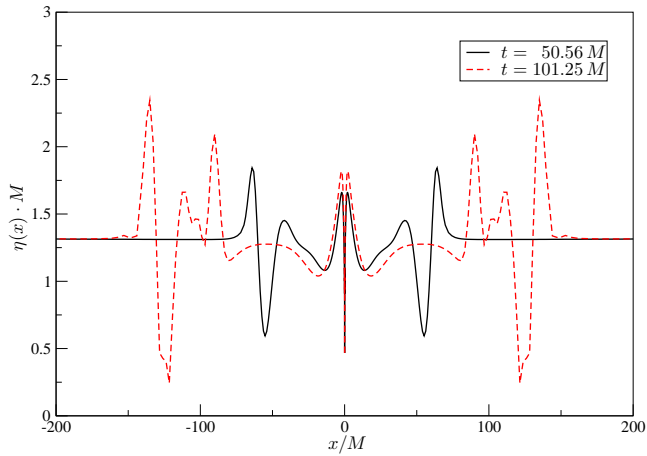}
  \caption{
  Form of $\er$ for a single spinning puncture sitting at $x=0$ using
  Eq.~(\ref{eq:etapsimm}) after simulation time $t=50.56\,M$ (solid black line)
  and $t=101.25\,M$(dashed  red line) over $x-$direction.
  }
  \label{fig:kbh_eta}
\end{figure}

To illuminate the origin of the disturbances in $\er$, we looked at
the development of $\er$ in simulations of a single, non-spinning
puncture, and a single, spinning puncture ($S_z/M^2 = 0.25$).  The
result for the spinning case is plotted in Fig.~\ref{fig:kbh_eta} at
two different times over the $x$-axis.  Again, we see a pulse
traveling outwards. Only this time, it does not leave much noise on
the grid.  The fact that this pulse travels at a speed which is
roughly $1.39$ (in our geometric units where $c=G=1$) in both the
spinning and non-spinning scenario indicates that it is related to the
gauge modes traveling at speed $\sqrt{2}$ in the asymptotic region
where $\alpha\simeq1$ (see \cite{GunGar06} and \cite{AlcBruDie02} for
a discussion of gauge speeds).  In contrast to gauge pulses in the
lapse, $\alpha$, or shift vector, $\beta^i$, the pulse in $\eta(x)$ is
amplified as it walks out. We found the same result in the single
puncture simulation without spin.  We believe the reason for this
behavior is that as the distance to the puncture increases, the
conformal factor, $\psi$, gets closer to unity. Therefore, the
denominator in Eq.~(\ref{eq:etapsimm}) approaches zero, and the gauge
disturbances in the derivatives of $\psi$ are magnified.  We further
observed reflections at the refinement boundaries as this pulse passes
through them. This may explain the fluctuations in $\eta(x)$ shown in
Fig.~\ref{fig:oldeta_waves}.  While one could continue to fine-tune a
formula dependent on the conformal factor to deal with these problems,
we looked in a different direction to determine the form of the
damping parameter.

\subsection{Formulas for $\eta$ with explicit dependence on the
  position and mass of the punctures}
Since we always know the location of a puncture, and we know
what its associated mass, we chose a form of damping that
uses this local information throughout the domain.  To address the
demands and concerns discussed in Section~\ref{sec:Motivation} and
\ref{sec:Prev}, we designed two position-dependent forms of
$\eta$. The two forms we tested are
\begin{equation}
  \er = A + \frac{C_1}{1 + w_1\,(\hat{r}_1^2)^n} + \frac{C_2}{1 +
    w_2\,(\hat{r}_2^2)^n},
  \label{eq:etaS6}
\end{equation}
and
\begin{equation}
    \er = A + C_1e^{-w_1\,(\hat{r}_1^2)^n} + C_2e^{-w_2\,(\hat{r}_2^2)^n}.
    \label{eq:etaS5}
\end{equation}
In Eqs.~(\ref{eq:etaS6}) and (\ref{eq:etaS5}), $w_1$ and $w_2$ are
required to be positive, unitless parameters which can be chosen to
change the width of the functions. The power $n$ is taken to be a
positive integer which determines the fall-off rate.  The constants
$A$, $C_1$, and $C_2$ are then chosen to provide the desired values of
$\eta$ at the punctures, and at at infinity.  Lastly, $\hat{r}_1$ and
$\hat{r}_2$ are defined as $\hat{r}_i =
\frac{|\vec{r}_i-\vec{r}|}{|\vec{r}_1-\vec{r}_2|}$, where $i$ is either
one or two, and $\vec{r}_i$ is the position of the $i$'th black hole.

The definition of $\hat{r}_i$ is chosen to naturally scale the
fall-off to the separation of the black holes.  $w_1$, $w_2$, and $n$
can be chosen to change the overall fall-off.  Our work focuses on the
choice $w_1 = w_2 = w$ and $n=1$. Following \cite{MueBru09}, we
construct the damping factor to have units of inverse mass.  We choose
$A= 2/M_{tot}$, where $M_{tot} \equiv M_1 + M_2$ is defined as the sum
of the irreducible masses.  We then take $C_i = 1/M_i - A$.  It is
then evident that both Eqs.~(\ref{eq:etaS6}) and (\ref{eq:etaS5}) will
give a constant value of $\eta = 2/M_{tot}$ in the equal mass case.

We designed the two formulas for $\eta$ in order to test the value of
using fundamentally different functions.  In our simulations, we found
little noticeable difference in the application of one compared to the other.
In the absence of such a difference, it becomes more beneficial to use
Eq.~(\ref{eq:etaS6}), as Gaussians are computationally more expensive.
It should be pointed out that Eq.~(\ref{eq:etaS6}) is very similar to
Eq.~(13) suggested in \cite{Sch10}, and we believe the following
results are very similar to what would be found using that form for
the damping. Going into the present work, we have no ansatz which might
suggest these forms of damping yield wave forms which are any better
than the use of any previous form of $\eta$. However, as will be seen
in the results sections, the waveforms we get from unequal mass
binaries show noticeable improvement over the constant $\eta$ case.

\section{Results}\label{sec:Results}

For data analysis purposes, we are mainly interested in the properties
of the emitted gravitational waves of the black hole binary systems
under study. Hence, it is important to check how the changes in the
gauge alter the extracted waves.  In the context of gravitational wave
extraction, $\Psi_4$ is only first order invariant under coordinate
transformations.  In addition, we have to chose an extraction radius
$r_{\rm ex}$ for the computation of modes, which is also coordinate
dependent. Although the last point can be partly addressed by
extrapolation of $r_{\rm ex}\rightarrow\infty$, it is a priori not
clear how much a change of coordinates affects the gravitational
waves. Furthermore, a change of coordinates implies an effective
change of the numerical resolution, and for practical purposes we have
to ask how much waveforms differ at a given finite resolution.

\subsection{Waveform comparison using formula (\ref{eq:etaS6})}
\label{sec:resultsS6}

The results in the following section refers to the use of Eq.~(\ref{eq:etaS6}).
We compare numerical simulations using three different grid
configurations, which correspond to three different resolutions.  In
the terminology of~\cite{BruGonHan06}, the grid set-ups are
$\phi[5\times64:7\times128:5]$, $\phi[5\times72:7\times144:5]$, and
$\phi[5\times80:7\times160:5]$, which corresponds to resolutions on
the finest grids of $3\,M/320$ ($N=64$), $M/120$ ($N=72$) and
$3\,M/400$ ($N=80$), respectively.  When referring to results from
different resolutions, we will from here on use the number of grid
points on the finest grid, $N$, to distinguish between them.  In this
subsection, we use $w_1=w_2=12$ and $n=1$ in Eq.~(\ref{eq:etaS6}).  As
test system we use an unequal mass black hole binary with mass ratio
$m_2/m_1=4$ and an initial separation of $D=5\,M$ without spins in
quasi-circular orbits.

\begin{figure}
  \centering
  \includegraphics[width=\linewidth,clip]{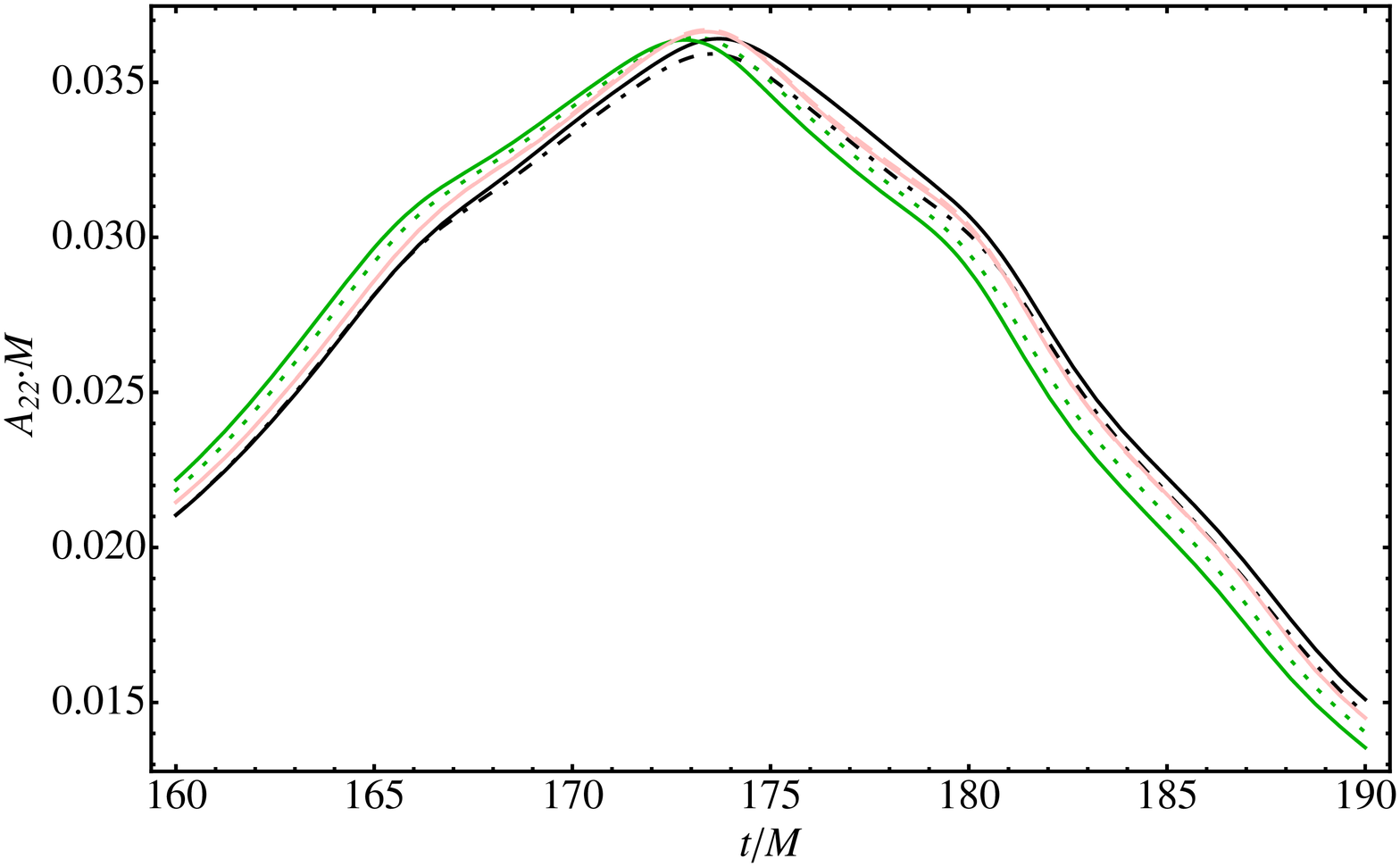}
  \put(-187,20){\includegraphics[width=0.47\linewidth,clip]{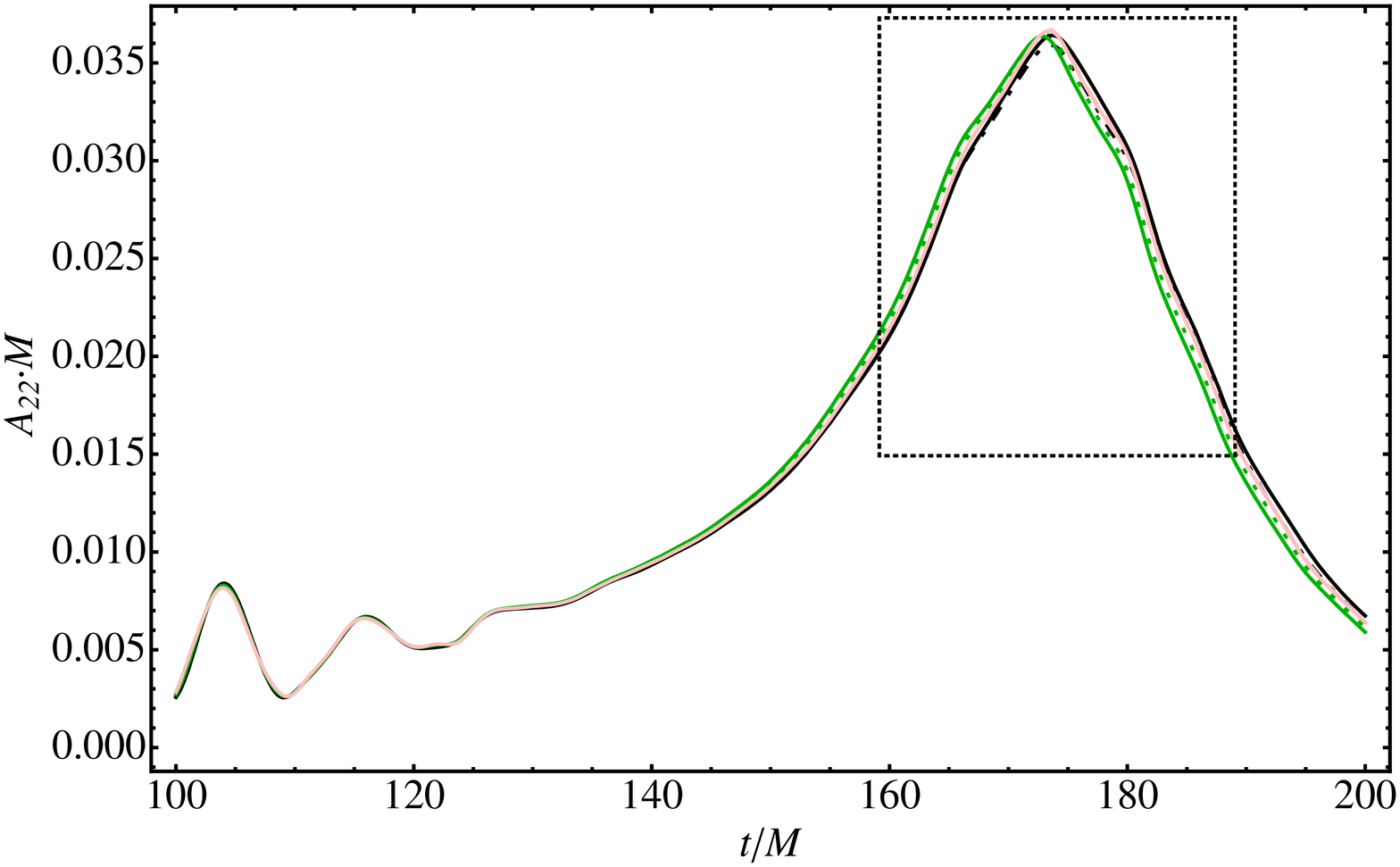}}
\put(-50,105){\includegraphics[width=0.2\linewidth,clip]{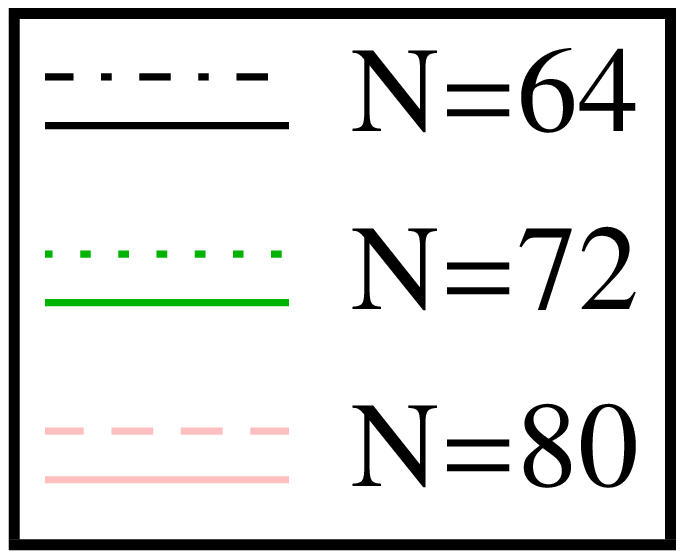}}
  \caption{Amplitude of the 22-mode of $\Psi_4$ of a binary with mass ratio 4:1
  and initial separation $D=5\,M$. The different colors correspond to three
  different resolutions according to the grid setup described in the text. The
  solid lines are results for $\eta=2/M$, the dashed, dotted and dot-dashed
  ones are for $\er$ (Eq.~(\ref{eq:etaS6})). The inset shows the simulation from
  shortly after the junk radiation passed, in the main plot we zoom into the
  region of highest amplitude (near the merger).
  }
  \label{fig:um4Amp}
\end{figure}

For orientation, Fig.~\ref{fig:um4Amp} shows the amplitude of the
22-mode, $A_{22}$, computed with the standard gauge $\eta=2/M$
(displayed as solid lines) and with the new $\er$ using
Eq.~(\ref{eq:etaS6}) (displayed as non-solid lines). The three
different colors correspond to the three resolutions. The inset shows
a larger time range of the simulation, while the main plot
concentrates on the time frame around merger.  The plot gives a course
view of the closeness of the results we obtain with standard and new
gauges.

\begin{figure}
  \centering
  \includegraphics[width=\linewidth,clip]{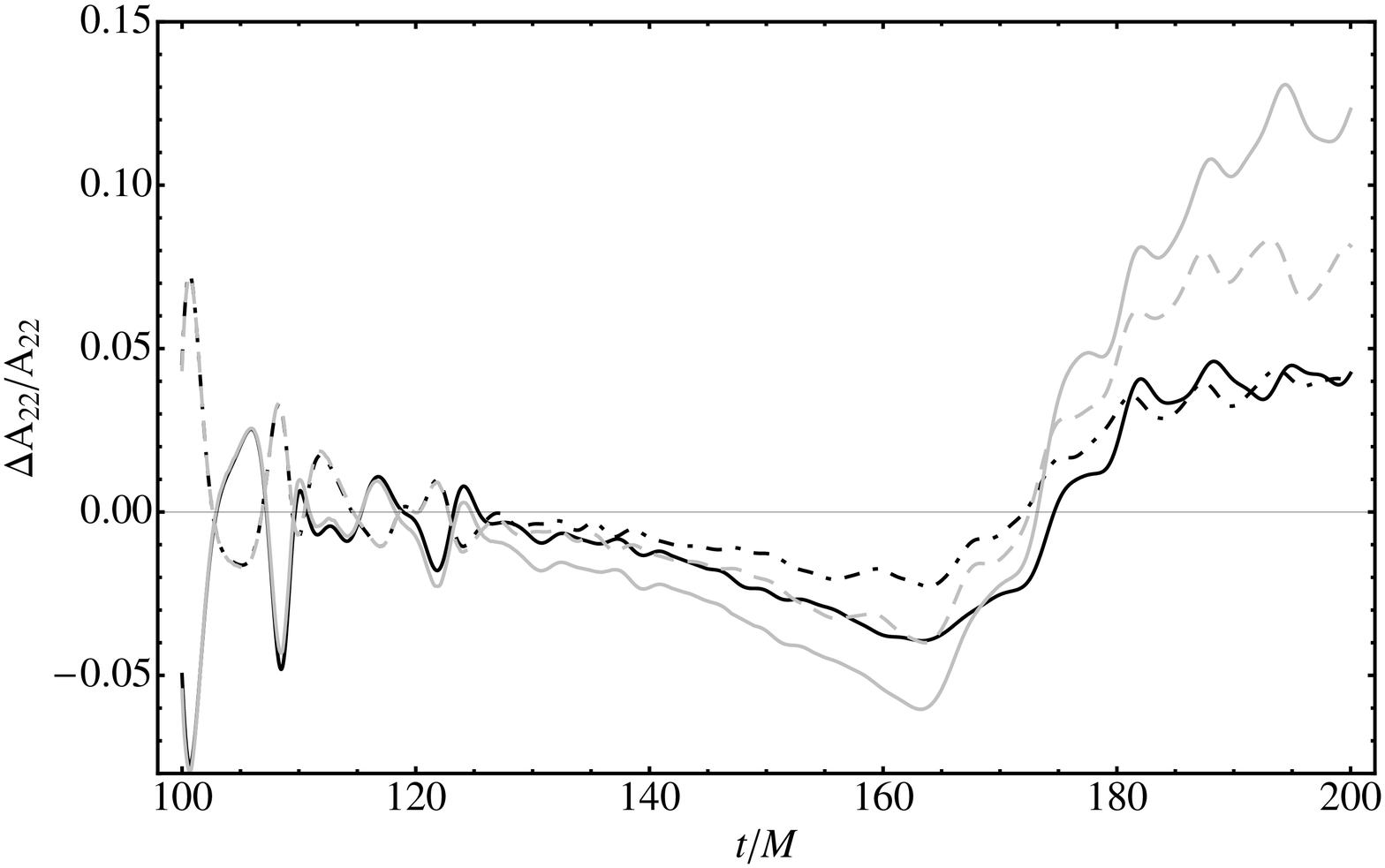}
\put(-200,100){\includegraphics[width=0.3\linewidth,clip]
{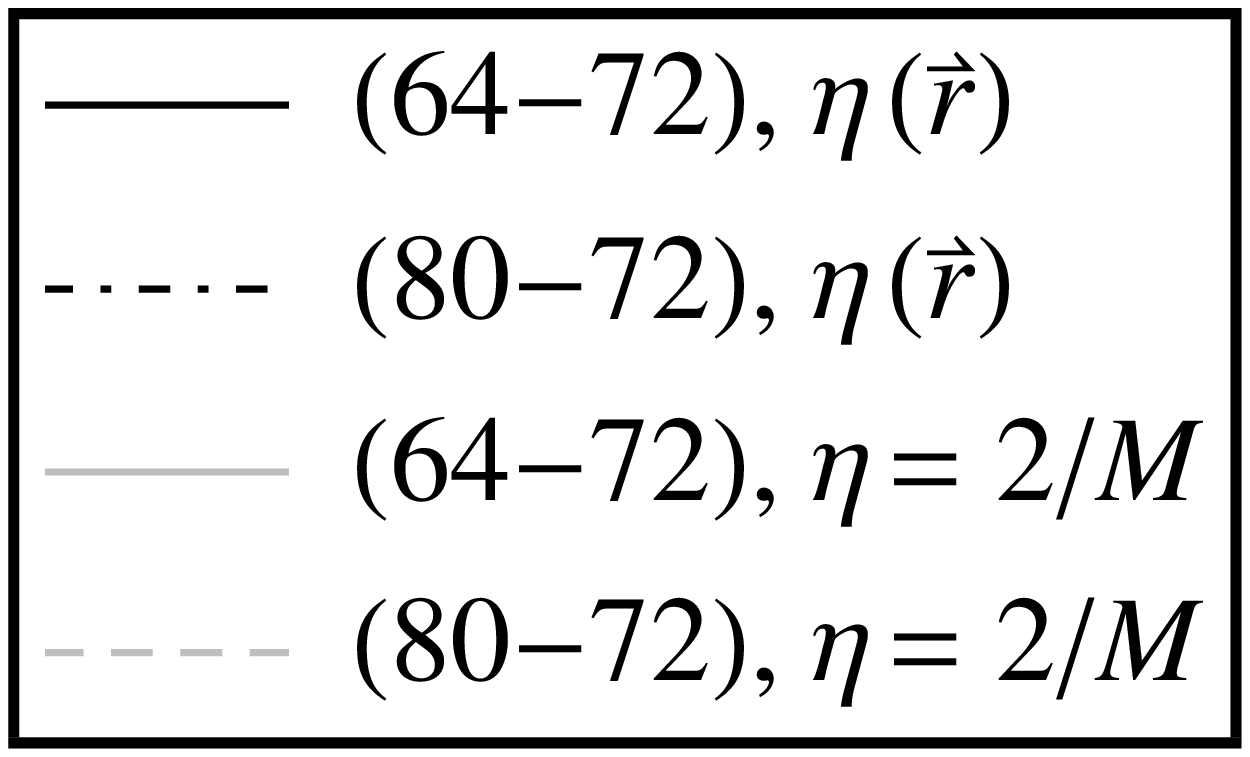}}
\caption{
  Relative differences of the amplitude of the 22-mode of $\Psi_4$
  between resolutions $N=64$ and $N=72$ (gray solid curve) as well as $N=72$ and
  $N=80$ (gray dashed curve) when using  $\eta=2/M$. 
  The same for $\er$ (Eq.~(\ref{eq:etaS6})) between $N=64$ and $N=72$ (black
  solid curve) and $N=72$ and $N=80$ (black dot-dashed curve).  The physical
  situation is the 
  same as in Fig.~\ref{fig:um4Amp}. The maximum differences are above $10\%$,
  comparing low and medium resolution of the constant
  $\eta$ simulations (gray solid line).
  }
  \label{fig:um4S6AmpDevRes2}
\end{figure}

In Fig.~\ref{fig:um4S6AmpDevRes2}, we plot the relative differences
between the amplitudes at low and medium (solid lines), and medium and
high resolution (non-solid lines) obtained with $\eta=2/M$ (light gray
lines) as well as $\er$ (Eq.~(\ref{eq:etaS6})) (black lines).  Here,
we find the maximum error between the low and medium resolution of the
series using $\eta=2/M$ amounts to about $12\%$ (solid gray
curve). Between medium and high resolution (dashed gray curve), we
find a smaller relative error, but it still goes up to $7\%$ at the end
of the simulation.  Employing Eq.~(\ref{eq:etaS6}), the maximum
amplitude error between low and medium resolution (solid black line)
is only about $4\%$, and therefore even smaller than the error between
medium and high resolution for the constant damping case. Between
medium and high resolution, the relative amplitude differences for
Eq.~(\ref{eq:etaS6}) are in general smaller than the ones between low
and medium resolution, although the maximum error is comparable to it
(dot-dashed black line).

\begin{figure}
  \centering
  \includegraphics[width=\linewidth]{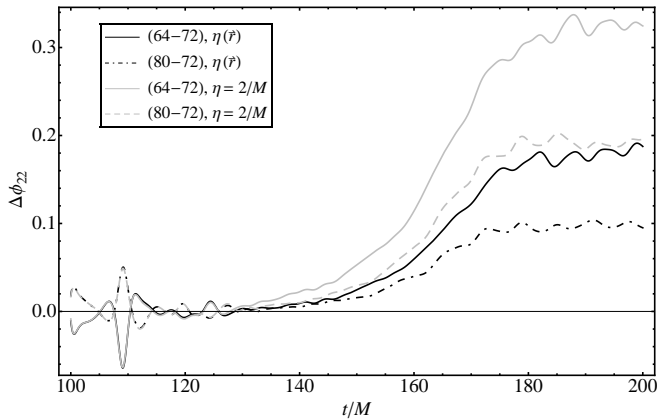}
\put(-215,105){\includegraphics[width=0.3\linewidth]
{legend_2DiffResNN.eps}}
  \caption{
  Phase differences between lowest and medium resolution for the series using
  $\eta=2/M$ (solid gray line) and $\er$ (Eq.~(\ref{eq:etaS6})) (solid black
  line)  as well as between medium and high resolution for $\eta=2/M$
  (dashed gray line) and for $\er$ (Eq.~(\ref{eq:etaS6})) (dot-dashed black
  line). The   physical situation is the one of Fig.~\ref{fig:um4Amp}. 
  }
  \label{fig:um4PhaseDevRes}
\end{figure}

We repeat the previous analysis for the phase of the 22-mode,
$\phi_{22}$.  Again, we compare the errors between resolutions in a
fixed gauge.  Figure~\ref{fig:um4PhaseDevRes} shows that the error
between lowest and medium resolution using $\eta=2/M$ (solid gray
line) grows up to about 0.31 radians.  For the differences between
medium and high resolution (dashed line) we find a maximal error of
0.2 radians for $\eta=2/M$.  For $\er$ following Eq.~(\ref{eq:etaS6}),
the phase error between low and medium resolution is only about 0.19
radians (solid black line) and decreases to 0.1 radians between medium
and high resolution (dot-dashed line).  Again, employing the position
dependent form of $\eta$, Eq.~(\ref{eq:etaS6}), the error between
lowest and medium resolution is lower than the one we obtain for
constant $\eta$ between medium and high resolution.  The results for
amplitude and phase error suggest that we can achieve the same
accuracy with less computational resources using a position-dependent
$\er$.


\subsection{Waveform comparison using formula (\ref{eq:etaS5})}
\label{sec:resultsS5}
We repeated the analysis of Sec.~\ref{sec:resultsS6} with the
waveforms we obtain using Eq.~(\ref{eq:etaS5}) (with $w_1=w_2=12$ and
$n=1$).  We use the same initial conditions (mass ratio $4:1$,
$D=5\,M$, no spins), and compare the amplitudes and phases of the
$22$-mode of $\Psi_4$ with the results of the $\eta=2/M$-runs. The
grid configurations remain the same.

\begin{figure}
  \centering
  \includegraphics[width=\linewidth]{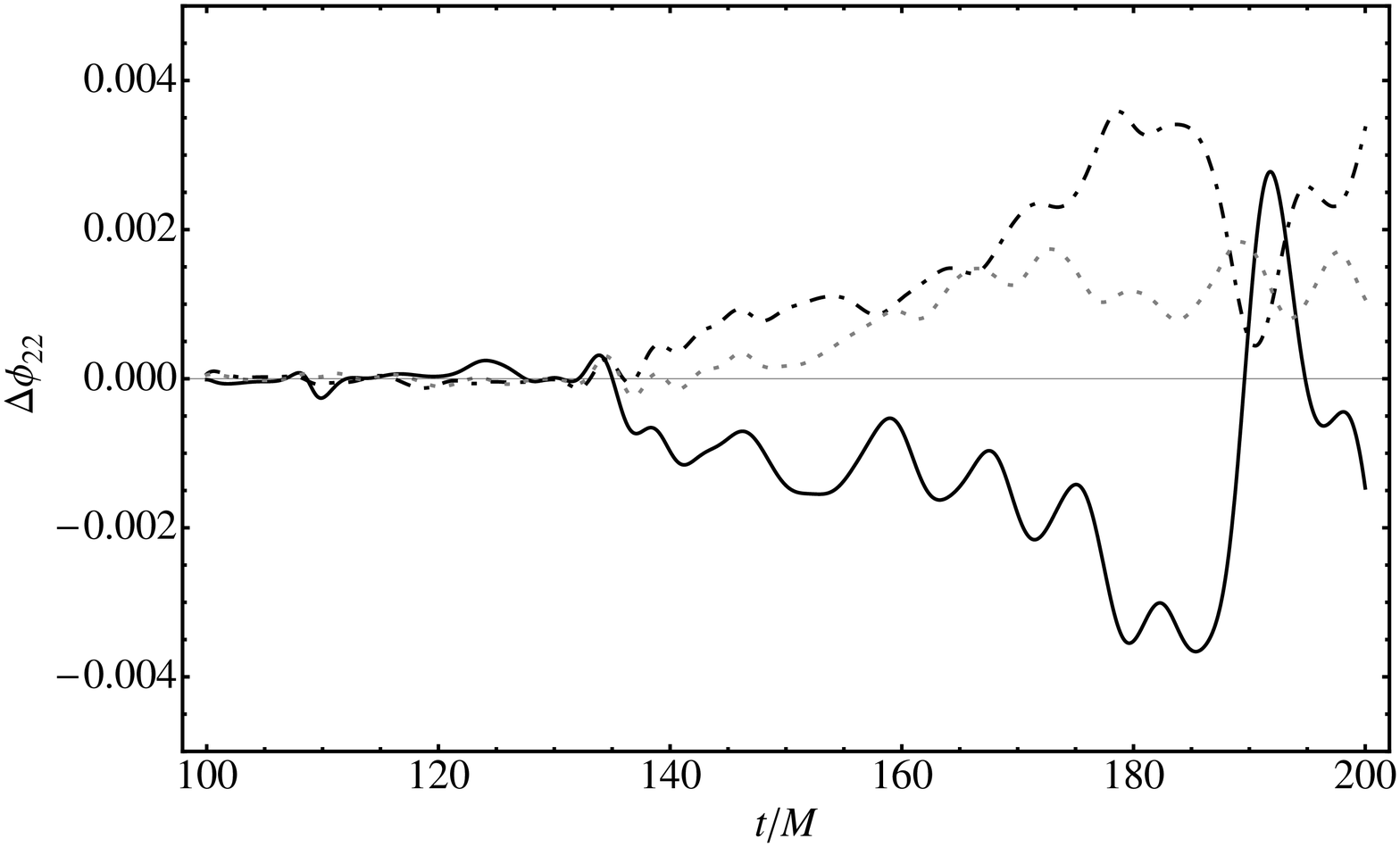}
\put(-215,100){\includegraphics[width=0.4\linewidth]
{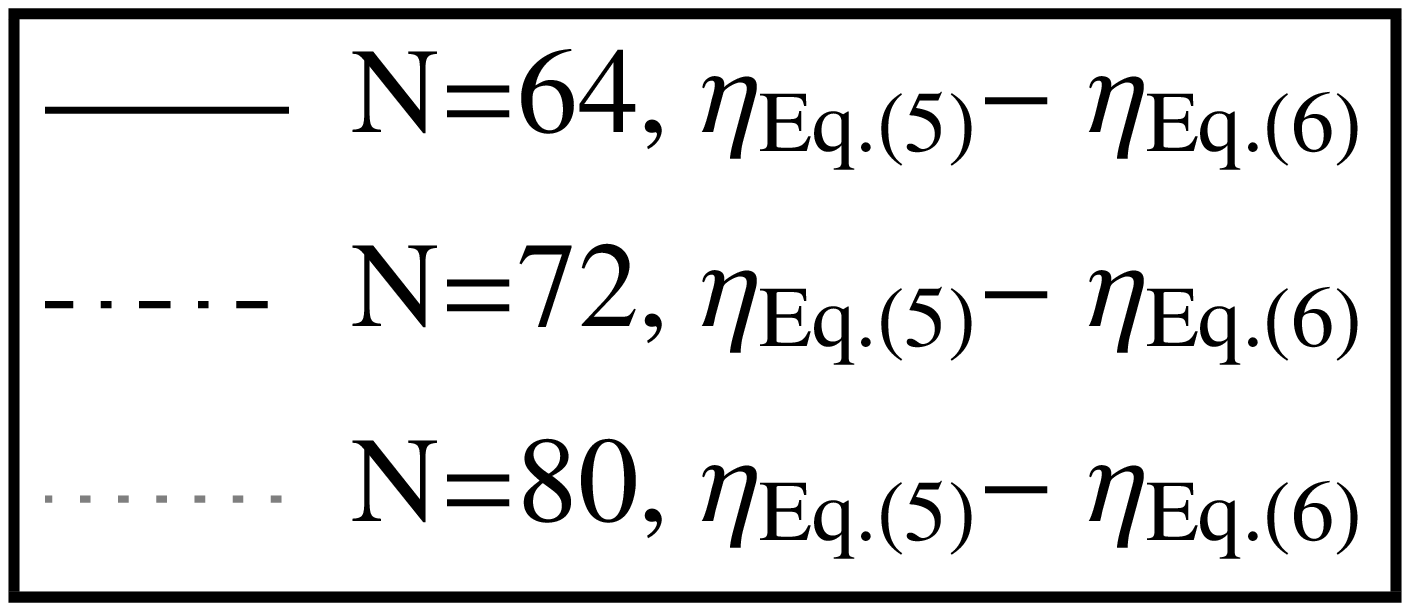}}
  \caption{
  Phase differences between waveforms obtained with Eq.~(\ref{eq:etaS6}) and
  Eq.~(\ref{eq:etaS5}) in three different resolutions (solid, dashed,
  dotted-dashed lines) for mass ratio 4:1, $D=5\,M$. 
  }
  \label{fig:um4PhaseDev_S5_S6}
\end{figure}

\begin{figure}
  \centering
  \includegraphics[width=\linewidth]{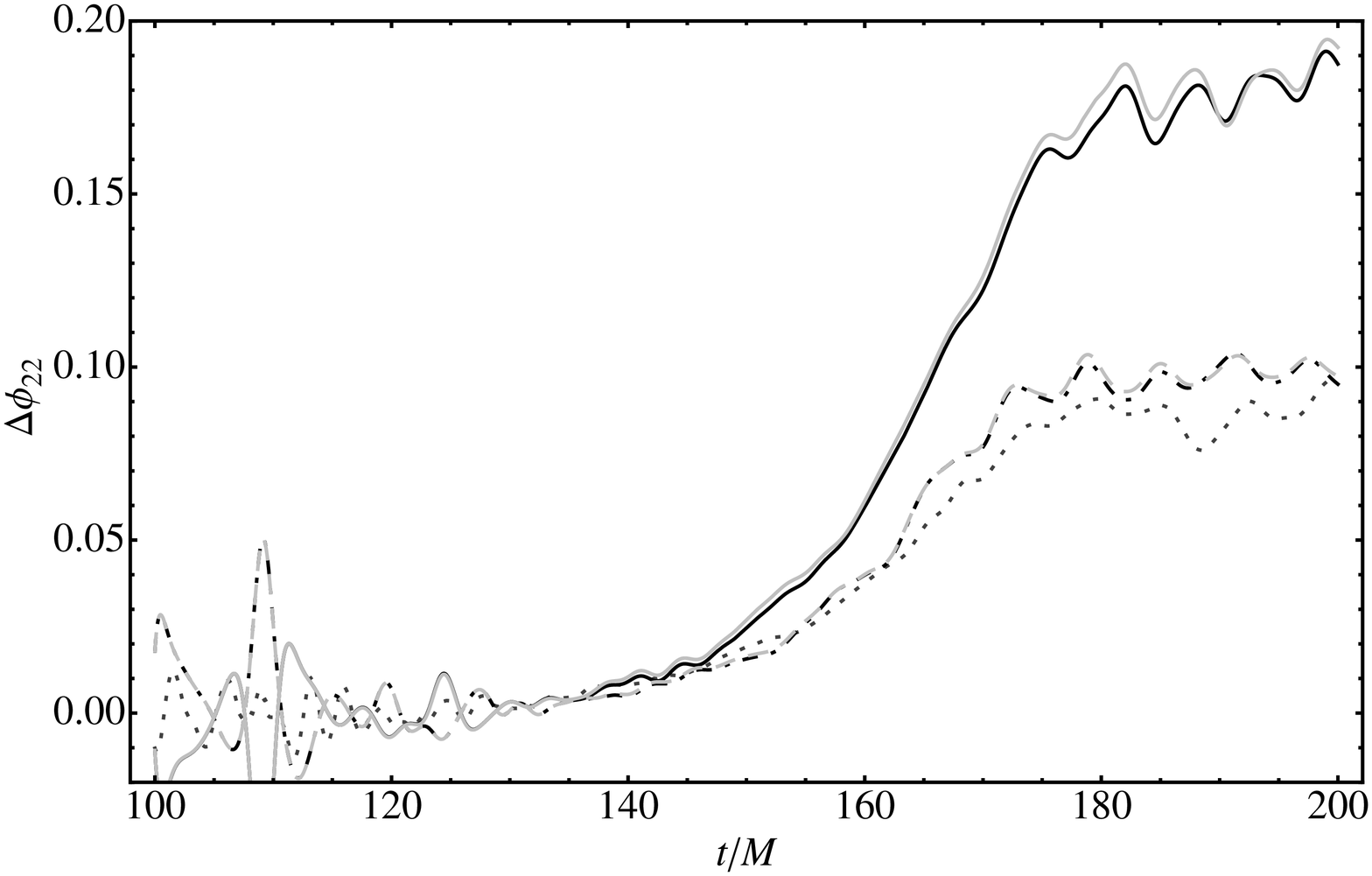}
\put(-215,80){\includegraphics[width=0.3\linewidth]
{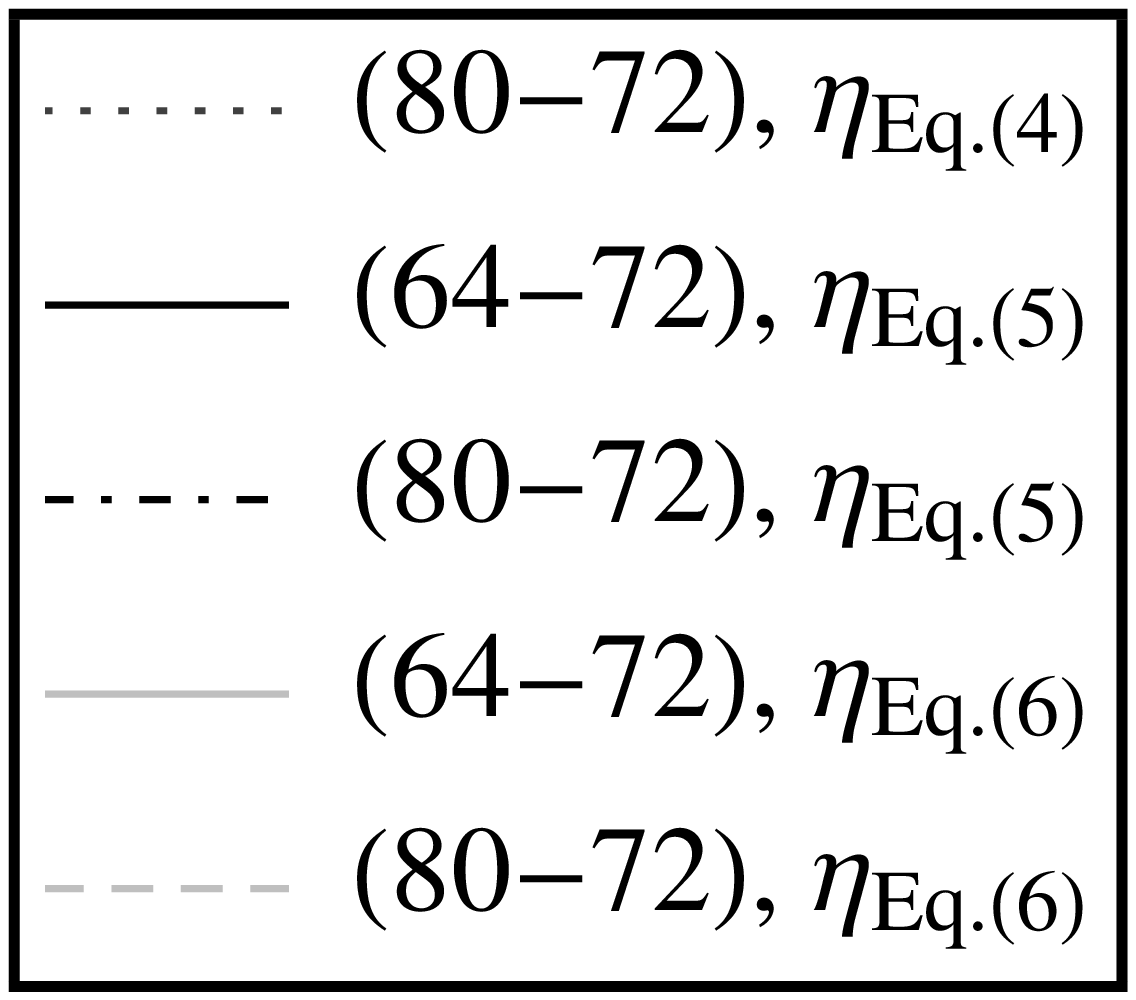}}
\caption{ Phase difference between waveforms at low and medium
  resolution (solid lines) and medium and high resolution
  (dotted-dashed and dashed line) using either Eq.~(\ref{eq:etaS6})
  (black lines) or Eq.~(\ref{eq:etaS5}) (gray lines) for mass ratio
  4:1, $D=5\,M$. For comparison, we also show the phase difference
  obtained with Eq.~(\ref{eq:etapsimm}) between medium and high
  resolution (dotted line).  }
  \label{fig:um4PhaseDevResS5S6Psim2}
\end{figure}

The results are very similar to the ones we obtained in
Figs.~\ref{fig:um4S6AmpDevRes2} and \ref{fig:um4PhaseDevRes}, and we
therefore do not show them here.  Although Eqs.~(\ref{eq:etaS6}) and
(\ref{eq:etaS5}) result in different shapes for $\er$, $\Psi_4^{22}$
is very similar.  Therefore, the comparison to $\eta=2/M$ naturally
gives very similar results, too.  The phase differences between results
from Eqs.~(\ref{eq:etaS6}) and (\ref{eq:etaS5}) at a given resolution are
shown in Fig.~\ref{fig:um4PhaseDev_S5_S6}. These are, with a maximum
phase error of 0.004 radians, very small compared to the phase errors
between resolutions, which, at minimum, are about 0.1 radian (see
Fig.~\ref{fig:um4PhaseDevRes}).
Fig.~\ref{fig:um4PhaseDevResS5S6Psim2} compares the phase error
between low and medium (solid lines), and medium and high resolution
(dotted-dashed and dashed line) of Eq.~(\ref{eq:etaS5}) (gray) to the
ones of Eq.~(\ref{eq:etaS6}) (black). For comparison, the error
between medium and high resolution is also plotted for
Eq.~(\ref{eq:etapsimm}) in this figure (dotted line). The plot
indicates that the errors between resolutions are in good agreement
for the different position dependent formulas of $\eta$.


\subsection{Behavior of the shift vector}\label{sec:results_betax}
\begin{figure}
  \centering
  \includegraphics[width=\linewidth,clip]{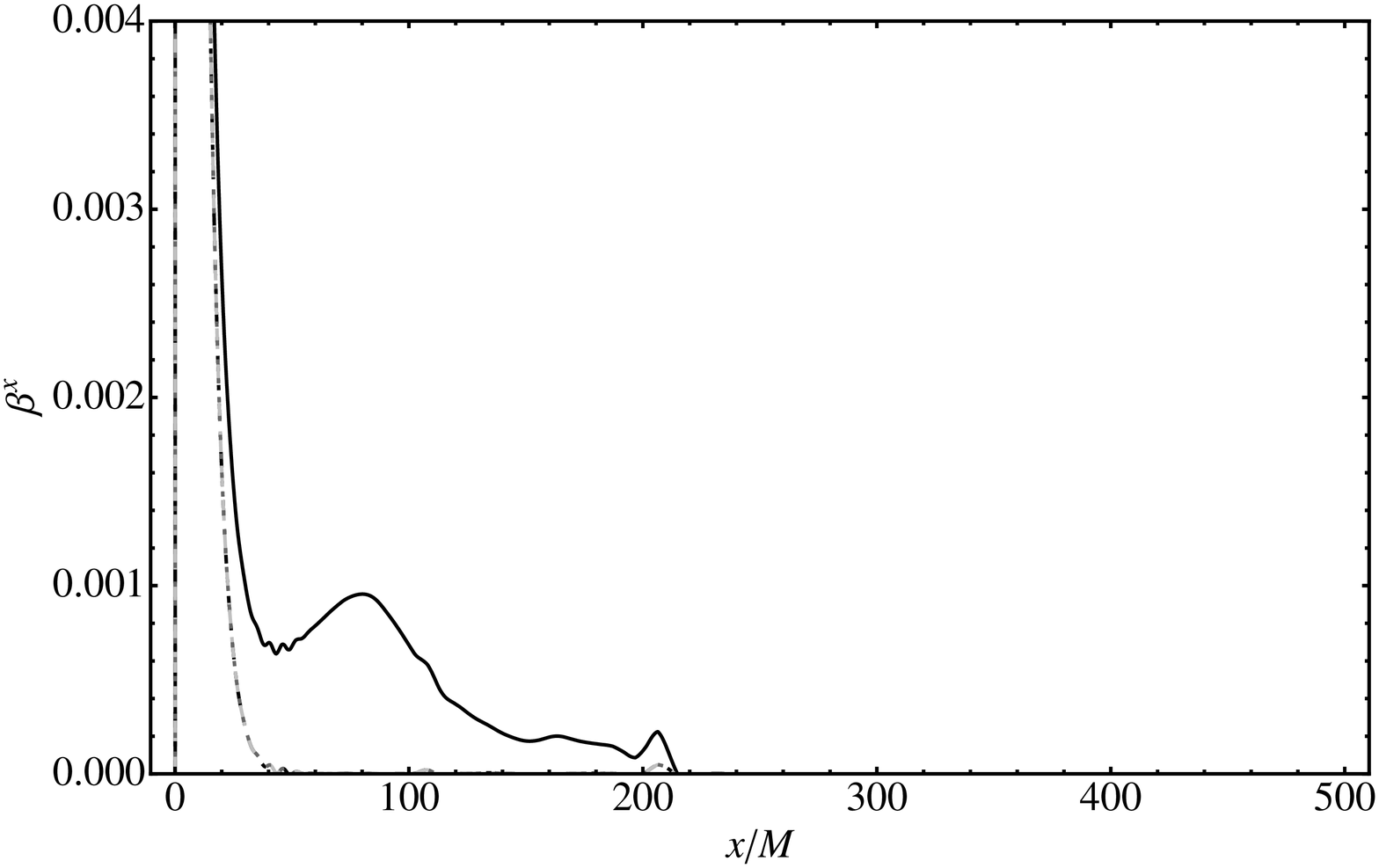}
\put(-90,80){\includegraphics[width=0.2\linewidth,clip]{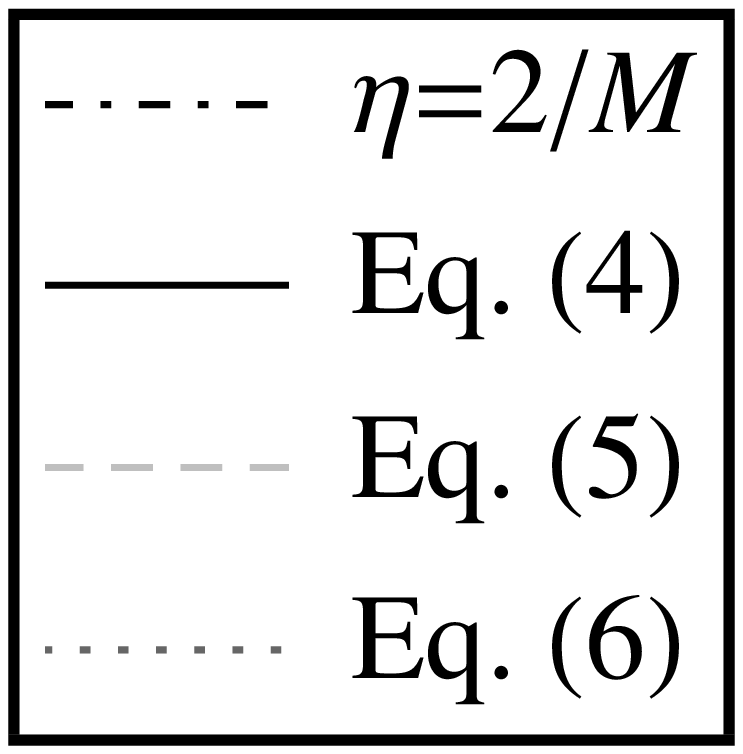}
}
  \caption{
  $x$-component of the shift vector in $x$-direction after $160\,M$ of
  evolution of the system with mass ratio $4:1$ and $D=10\,M$.  The black,
  dot-dashed line refers to the use of a constant damping $\eta$,
  while the black,
  solid line uses Eq.~(\ref{eq:etapsimm}). The gray, dashed line is for the use
of
  Eq.~(\ref{eq:etaS5}) and the gray, dotted one for Eq.~(\ref{eq:etaS6}). Except
for
  the constant $\eta$ (black, dot-dashed line), the results in this plot are
  indistinguishable. 
 }
  \label{fig:um4betax}
\end{figure}
In \cite{MueBru09}, we found an unusual behavior of the shift vector.
This is illustrated in Fig.~\ref{fig:um4betax}, where we plot the
$x$-component of the shift, $\beta^x$, in the $x$-direction after
$160\,M$ of evolution (this means approximately $80\,M$ after merger)
for all four versions of the damping constant we used for comparison
in this paper before, and for the same binary configuration as the one
used in Secs.~\ref{sec:resultsS6} and \ref{sec:resultsS5}. Like in
\cite{MueBru09}, we find that using Eq.~(\ref{eq:etapsimm}) results in
a shift which falls off to zero too slowly towards the outer boundary,
and which develops a ``bump'' (black, solid line), while the constant
damping case (black, dot-dashed line) falls off to zero quickly.
Employing Eqs.~(\ref{eq:etaS6}) or (\ref{eq:etaS5}) avoids this
undesirable feature. After merger, the shift falls off to zero when
going away from the punctures as it does in the constant damping case
(gray dashed and dotted lines).  Using Eq.~(\ref{eq:etaS6}) or
(\ref{eq:etaS5}) prevents unwanted coordinate drifts at the end of the
simulations.

\section{Discussion}\label{sec:Discussion}

In this work, we examined the role that the damping factor, $\eta$, plays
in the evolution of the shift when using the gamma
driver. In particular, we examined the range of values allowed in
various evolutions, and what effects showed up because of the value
chosen.  We then designed a form of $\eta$ for the evolution of binary
black holes which provides appropriate values both near the individual
punctures and far away from them with a smooth transition in between.

In Sec.~\ref{sec:Results}, we directly examined the waveforms for the
case using Eq.~(\ref{eq:etaS6}), where $w_1=w_2=12$ and $n=1$.  While
the form of $\eta$ is predictable, and can be easily adjusted for
stability, we also saw that the waveforms produced using this
definition showed less deviation with increasing resolution than using
a constant $\eta$. When examining the waveforms produced using
Eq.~(\ref{eq:etaS5}), we found similar results.  In the absence of a
noticeable difference in the quality of the waveforms,
Eq.~(\ref{eq:etaS6}) is computationally cheaper, and, as such, is our
preferred definition for the damping.

\begin{figure}
    \centering
    \includegraphics[width=\linewidth,clip]{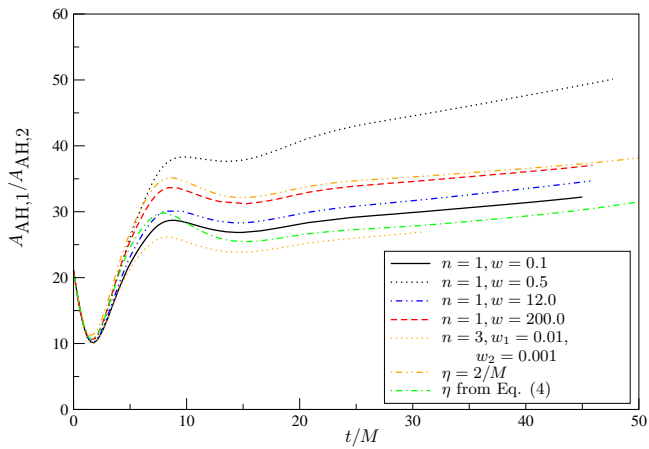}
    \caption{Shown is the time
    dependence of the ratio between the coordinate
    areas of the apparent  horizons of both black holes in a simulation with
    mass ratio $4:1$ with initial separation  $D=5\,M$.
    The black, blue and red lines use $\er$, Eq.~(\ref{eq:etaS6}) with varying
    values of the width parameter $w$.
    The orange line (dash-dot-dot) uses the constant damping $\eta=2/M$ and
    the green (dash-dot) one refers to the result of \cite{MueBru09} with
    Eq.~(\ref{eq:etapsimm}).
    Using Eq.~(\ref{eq:etaS6}), the coordinate areas can be varied with
    respect to each other depending on the choice of $w$. A ratio of 1 means
    the black holes have the same size on the numerical grid.}
    \label{fig:AH_coordArea}
  \end{figure}

We have already pointed out a certain freedom to pick parameters in
Eqs.~(\ref{eq:etaS6}) and (\ref{eq:etaS5}). We did perform some
experimentation along this line where we varied $w=w_1=w_2$ to see if
we could get a useful effect of the coordinate size of the apparent
horizons on the numerical grid.  In~\cite{BruGonHan06,HerShoLag06}, it
was noticed that the damping coefficient affects the coordinate
location of the apparent horizon, and therefore the resolution of the
black hole on the numerical grid.  Fig.~\ref{fig:AH_coordArea} plots
the ratio of the grid-area of larger apparent horizon to the smaller
apparent horizon as a function of time for $w$-values of $0.1$, $0.5$
and for $200$, all with $n=1$. Also plotted is the relative coordinate
size for the same binaries using a constant $\eta$ in dashed,
double-dotted line, and for using Eq.~(\ref{eq:etapsimm}) in a blue
dashed-dotted line. All the evolutions show an immediate dip, and then
increase in the grid-area ratio during the course of the
evolution. While a very low ratio was found using
Eq.~(\ref{eq:etapsimm}), the orange dotted line was later found for
the choices of $n=3$ with $w_1=0.01$ and $w_2=0.0001$ with
Eq.~(\ref{eq:etaS6}). Due to this freedom in the implementation of our
explicit formula for the damping, it may be possible to further reduce
the relative grid size of the black holes. This effect could be
important in easing the computational difficulty of running a
numerical simulation for unequal mass binaries.

Having a form of $\eta$ that leads to stable evolutions for any
mass-ratio is an important step towards the numerical evolution of
binary black holes in the intermediate mass-ratio. We believe the form
given in Eq.~(\ref{eq:etaS6}) provides such a damping factor at a
low computational cost, although the test results presented are
limited to mass ratio $4:1$.  We plan to examine larger mass ratios in
future work. The new method should allow binary simulations for mass
ratio $10:1$, or even $100:1$. It remains to be seen whether other
issues than the gauge are now the limiting factor for simulations at
large mass ratios.

\section*{Acknowledgments}
It is a pleasure to thank Zhoujian Cao and Erik Schnetter for discussions.
This work was supported in part by DFG grant SFB/Transregio~7
``Gravitational Wave Astronomy'' and the DLR (Deutsches Zentrum f\"ur Luft
und Raumfahrt). D. M. was additionally supported by the DFG Research
Training Group 1523 ``Quantum and Gravitational Fields''.
Computations were performed on the HLRB2 at LRZ Munich.

%%%%%%%%%%%%%%%%%%%%%%%%%%%%%%%%%%%%%%%%%%%%%%%%%%%%%%%%%%%%%%%%%%%%%%%%%%%%%%%%
\newpage
%\bibliography{refs}

\end{document}